%Paper: hep-th/9209071
%From: moreno%uimrl.dnet@uimrl7.mrl.uiuc.edu
%Date: Sat, 19 Sep 92 17:39:40 -0500

%%%%%%%%%%%%%%% USE PLAIN TEX (all macros included)

%%	JNL.TEX					Doug Eardley

\font\twelverm=cmr10 scaled 1200    \font\twelvei=cmmi10 scaled 1200
\font\twelvesy=cmsy10 scaled 1200   \font\twelveex=cmex10 scaled 1200
\font\twelvebf=cmbx10 scaled 1200   \font\twelvesl=cmsl10 scaled 1200
\font\twelvett=cmtt10 scaled 1200   \font\twelveit=cmti10 scaled 1200

\skewchar\twelvei='177   \skewchar\twelvesy='60

\def\twelvepoint{\normalbaselineskip=12.4pt plus 0.1pt minus 0.1pt
  \abovedisplayskip 12.4pt plus 3pt minus 9pt
  \belowdisplayskip 12.4pt plus 3pt minus 9pt
  \abovedisplayshortskip 0pt plus 3pt
  \belowdisplayshortskip 7.2pt plus 3pt minus 4pt
  \smallskipamount=3.6pt plus1.2pt minus1.2pt
  \medskipamount=7.2pt plus2.4pt minus2.4pt
  \bigskipamount=14.4pt plus4.8pt minus4.8pt
  \def\rm{\fam0\twelverm}          \def\it{\fam\itfam\twelveit}%
  \def\sl{\fam\slfam\twelvesl}     \def\bf{\fam\bffam\twelvebf}%
  \def\mit{\fam 1}                 \def\cal{\fam 2}%
  \def\tt{\twelvett}
  \textfont0=\twelverm   \scriptfont0=\tenrm   \scriptscriptfont0=\sevenrm
  \textfont1=\twelvei    \scriptfont1=\teni    \scriptscriptfont1=\seveni
  \textfont2=\twelvesy   \scriptfont2=\tensy   \scriptscriptfont2=\sevensy
  \textfont3=\twelveex   \scriptfont3=\twelveex  \scriptscriptfont3=\twelveex
  \textfont\itfam=\twelveit
  \textfont\slfam=\twelvesl
  \textfont\bffam=\twelvebf \scriptfont\bffam=\tenbf
  \scriptscriptfont\bffam=\sevenbf
  \normalbaselines\rm}

\def\beginlinemode{\endmode
  \begingroup\parskip=0pt \obeylines\def\\{\par}\def\endmode{\par\endgroup}}
\def\beginparmode{\endmode
  \begingroup \def\endmode{\par\endgroup}}
\let\endmode=\par
{\obeylines\gdef\
{}}
\def\singlespace{\baselineskip=\normalbaselineskip}

\def\oneandahalfspace{\baselineskip=\normalbaselineskip
  \multiply\baselineskip by 3 \divide\baselineskip by 2}
\def\doublespace{\baselineskip=\normalbaselineskip \multiply\baselineskip by 2}

\newcount\firstpageno
\firstpageno=2
\footline={\ifnum\pageno<\firstpageno{\hfil}\else{\hfil\twelverm\folio\hfil}\fi}
\def\toppageno{\global\footline={\hfil}\global\headline
  ={\ifnum\pageno<\firstpageno{\hfil}\else{\hfil\twelverm\folio\hfil}\fi}}
\let\rawfootnote=\footnote		% We must set the footnote style
\def\footnote#1#2{{\rm\singlespace\parindent=0pt\parskip=0pt
  \rawfootnote{#1}{#2\hfill\vrule height 0pt depth 6pt width 0pt}}}
\def\raggedcenter{\leftskip=4em plus 12em \rightskip=\leftskip
  \parindent=0pt \parfillskip=0pt \spaceskip=.3333em \xspaceskip=.5em
  \pretolerance=9999 \tolerance=9999
  \hyphenpenalty=9999 \exhyphenpenalty=9999 }
\def\dateline{\rightline{\ifcase\month\or
  January\or February\or March\or April\or May\or June\or
  July\or August\or September\or October\or November\or December\fi
  \space\number\year}}
\def\received{\vskip 3pt plus 0.2fill
 \centerline{\sl (Received\space\ifcase\month\or
  January\or February\or March\or April\or May\or June\or
  July\or August\or September\or October\or November\or December\fi
  \qquad, \number\year)}}

\hsize=6.5truein
%\hoffset=1truein
\vsize=8.9truein
%\voffset=1truein
\parskip=\medskipamount
\def\\{\cr}
\twelvepoint		% selects twelvepoint fonts (cf. \tenpoint)
\doublespace		% selects double spacing for main part of paper (cf.
			%	\singlespace, \oneandahalfspace)
\overfullrule=0pt	% delete the nasty little black boxes for overfull box

  %  "Draft", Timestamp

	% Preprint number at upper right of title page

\def\title			%  Title on title page
  {\null\vskip 3pt plus 0.2fill
   \beginlinemode \doublespace \raggedcenter \bf}

\def\author			%  Author(s) name(s)  on title page
  {\vskip 3pt plus 0.2fill \beginlinemode
   \singlespace \raggedcenter}

\def\affil			% Affiliations (can intermix with \author)
  {\vskip 3pt plus 0.1fill \beginlinemode
   \oneandahalfspace \raggedcenter \sl}

\def\abstract			% Begin abstract
  {\vskip 3pt plus 0.3fill \beginparmode
   \oneandahalfspace ABSTRACT: }

\def\endtopmatter		% End title page, begin body of paper
  {\endpage			% 	This subsumes \body
   \body}

\def\body			% Begin text body;  can be used to end
  {\beginparmode}		% \title, \author, \affil, \abstract,
				% \reference, or \figurecaption modes

\def\head#1{			% Head;  NOTE enclose the text in {}
  \goodbreak\vskip 0.5truein	%  e.g., \head{I. Introduction}
  {\immediate\write16{#1}
   \raggedcenter \uppercase{#1}\par}
   \nobreak\vskip 0.25truein\nobreak}

\def\beneathrel#1\under#2{\mathrel{\mathop{#2}\limits_{#1}}}

\def\refto#1{$^{#1}$}		% For references in text as superscript

\def\references			% Begin references -- basic format is Phys Rev
  {\head{References}		% I.e., volume, page, year (space after commas).
   \beginparmode
   \frenchspacing \parindent=0pt \leftskip=1truecm
   \parskip=8pt plus 3pt \everypar{\hangindent=\parindent}}

\gdef\refis#1{\item{#1.\ }}			% Ref list numbers.

\gdef\journal#1, #2, #3, 1#4#5#6{		% Journal reference.  Comma sets
    {\sl #1~}{\bf #2}, #3 (1#4#5#6)}		% off: name, vol, page, year

\def\refstyleprnp{		% Input like pr, output like np!!
  \gdef\refto##1{ [##1]}			% Reference in text []
  \gdef\refis##1{\item{##1)\ }}			% Reference list numbers)
  \gdef\journal##1, ##2, ##3, 1##4##5##6{	% Journal reference
    {\sl ##1~}{\bf ##2~}(1##4##5##6) ##3}}

\def\prd{\journal Phys. Rev. D, }

\def\prl{\journal Phys. Rev. Lett., }

\def\rmp{\journal Rev. Mod. Phys., }

\def\cmp{\journal Comm. Math. Phys., }

\def\np{\journal Nucl. Phys., }

\def\pl{\journal Phys. Lett., }

\def\endreferences{\body}

\def\figurecaptions		% Begin figure captions
  {\endpage
   \beginparmode
   \head{Figure Captions}
}

\def\endpage			%  Eject a page
  {\vfill\eject}

\def\endpaper			%  Ways to say goodbye
  {\endmode\vfill\supereject}

\def\heading				% Heading
  {\vskip 0.5truein plus 0.1truein	% e.g., \heading I. NOTES \endheading
   \beginparmode \def\\{\par} \parskip=0pt \singlespace \raggedcenter}

\def\subheading				% Subheading
  {\vskip 0.25truein plus 0.1truein	% e.g., \subheading{A. The Problem}
   \beginlinemode \singlespace \parskip=0pt \def\\{\par}\raggedcenter}

\def\tag#1$${\eqno(#1)$$}

\def\align#1$${\eqalign{#1}$$}

\def\aligntag#1$${\gdef\tag##1\\{&(##1)\cr}\eqalignno{#1\\}$$
  \gdef\tag##1$${\eqno(##1)$$}}

\def\overset#1\to#2{{\mathop{#2}^{#1}}}
\def\underset#1\to#2{{\mathop{#2}_{#1}}}

\def\ref#1{Ref.~#1}			% 	for inline references
\def\Ref#1{Ref.~#1}			% 	ditto
\def\[#1]{[\cite{#1}]}
\def\cite#1{{#1}}

%%      reforder.tex

\catcode`@=11
\newcount\r@fcount \r@fcount=0
\newcount\r@fcurr
\immediate\newwrite\reffile
\newif\ifr@ffile\r@ffilefalse
\def\w@rnwrite#1{\ifr@ffile\immediate\write\reffile{#1}\fi\message{#1}}

\def\writer@f#1>>{}
\def\referencefile{%                      Stuff to write .REF file
  \r@ffiletrue\immediate\openout\reffile=\jobname.ref%
  \def\writer@f##1>>{\ifr@ffile\immediate\write\reffile%
    {\noexpand\refis{##1} = \csname r@fnum##1\endcsname = %
     \expandafter\expandafter\expandafter\strip@t\expandafter%
     \meaning\csname r@ftext\csname r@fnum##1\endcsname\endcsname}\fi}%
  \def\strip@t##1>>{}}

\def\citeall#1{\xdef#1##1{#1{\noexpand\cite{##1}}}}
\def\cite#1{\each@rg\citer@nge{#1}}     % Variable No. of args, separated by
% ","

\def\each@rg#1#2{{\let\thecsname=#1\expandafter\first@rg#2,\end,}}
\def\first@rg#1,{\thecsname{#1}\apply@rg}       % each@ag is a general purpose
\def\apply@rg#1,{\ifx\end#1\let\next=\relax%      variable no. of arg. macro.
\else,\thecsname{#1}\let\next=\apply@rg\fi\next}% args separated by commas

\def\citer@nge#1{\citedor@nge#1-\end-}  % Check for M-N range (M and N numbers)
\def\citer@ngeat#1\end-{#1}
\def\citedor@nge#1-#2-{\ifx\end#2\r@featspace#1 % Single argument
  \else\citel@@p{#1}{#2}\citer@ngeat\fi}        % M-N range of arguments
\def\citel@@p#1#2{\ifnum#1>#2{\errmessage{Reference range #1-#2\space is bad.}%
    \errhelp{If you cite a series of references by the notation M-N, then M and
    N must be integers, and N must be greater than or equal to M.}}\else%
 {\count0=#1\count1=#2\advance\count1
by1\relax\expandafter\r@fcite\the\count0,%

  \loop\advance\count0 by1\relax%         Loop from M to N
    \ifnum\count0<\count1,\expandafter\r@fcite\the\count0,%
  \repeat}\fi}

\def\r@featspace#1#2 {\r@fcite#1#2,}    % Eat spaces at beginning or end of arg
\def\r@fcite#1,{\ifuncit@d{#1}%           Cite individual reference
    \newr@f{#1}%
    \expandafter\gdef\csname r@ftext\number\r@fcount\endcsname%
                     {\message{Reference #1 to be supplied.}%
                      \writer@f#1>>#1 to be supplied.\par}%
 \fi%
 \csname r@fnum#1\endcsname}
\def\ifuncit@d#1{\expandafter\ifx\csname r@fnum#1\endcsname\relax}%
\def\newr@f#1{\global\advance\r@fcount by1%
    \expandafter\xdef\csname r@fnum#1\endcsname{\number\r@fcount}}

\let\r@fis=\refis                       % Save old \refis, redefine
\def\refis#1#2#3\par{\ifuncit@d{#1}%      Use two params #2 #3 to strip blank
   \newr@f{#1}%
   \w@rnwrite{Reference #1=\number\r@fcount\space is not cited up to now.}\fi%
  \expandafter\gdef\csname r@ftext\csname r@fnum#1\endcsname\endcsname%
  {\writer@f#1>>#2#3\par}}

\def\ignoreuncited{%   redefine \refis if ignoring uncited references
   \def\refis##1##2##3\par{\ifuncit@d{##1}%
     \else\expandafter\gdef\csname r@ftext\csname
r@fnum##1\endcsname\endcsname%

     {\writer@f##1>>##2##3\par}\fi}}

\def\r@ferr{\endreferences\errmessage{I was expecting to see
\noexpand\endreferences before now;  I have inserted it here.}}
\let\r@ferences=\references
\def\references{\r@ferences\def\endmode{\r@ferr\par\endgroup}}

\let\endr@ferences=\endreferences
\def\endreferences{\r@fcurr=0%            Save old \endreferences, redefine
  {\loop\ifnum\r@fcurr<\r@fcount%         Loop over refnum and produce text
    \advance\r@fcurr by 1\relax\expandafter\r@fis\expandafter{\number\r@fcurr}%
    \csname r@ftext\number\r@fcurr\endcsname%
  \repeat}\gdef\r@ferr{}\endr@ferences}

% Save old \endpaper, redefine it to write parting message.

\let\r@fend=\endpaper\gdef\endpaper{\ifr@ffile
\immediate\write16{Cross References written on []\jobname.REF.}\fi\r@fend}

\catcode`@=12

\citeall\refto          % These macros will generate citations
\citeall\ref            %
\citeall\Ref            %

\refstyleprnp

\def\slp{{\raise.15ex\hbox{$/$}\kern-.57em\hbox{$\partial$}}}
\def\lnA{\raise.15ex\hbox{$/$}\kern-.57em\hbox{$A$}}
\def\slD{\raise.15ex\hbox{$/$}\kern-.57em\hbox{$D$}}
\def\lncA{\raise.15ex\hbox{$/$}\kern-.57em\hbox{$\cal A$}}
\def\ze{{\zeta}}
\def\th{{\theta}}
\def\cA{{\cal A}}
\def\cD{{\cal D}}
\def\cL{{\cal L}}
\def\cR{{\cal R}}
\def\cS{{\cal S}}
\def\cZ{{\cal Z}}
\def\sun{${\hat {su}}$({\tenit N})}
\def\ee{\hat {\bf e}}
\def\pz{\partial_z}
\def\jz{{\bf j}_z^{N-1}}
\def\rh{{\bf \rho}}
\def\total{$\underline{\hbox{\rm total}}$}
\def\ie{{\it i.e.,\ }}
\def\via{{\it via\ }}

\title COSET MODELS IN MONOPOLE BACKGROUNDS

\author ENRIQUE F. MORENO

\affil Department of Physics
       University of Illinois at Urbana-Champaign
       1110 W. Green St., Urbana, IL 61801, USA.

\abstract {We study a fermionic coset model $G/H$ when the subgroup $H$ is not
simply connected. We show that even when the fermionic zero modes impose
selection rules which alters the values of the correlators, the Virasoro
central charge of the theory results independent of the topological sector.}

\endtopmatter

\head{Section I: Introduction}

The $G/H$ coset construction of conformal algebras has proven to be extremely
useful in the classification of conformal field theories\refto{GKO}. In
particular, it provides a method to obtain new representations of the Virasoro
algebra by decomposing the representations of the original group $G$ Kac Moody
algebras into representations of a subgroup $H$ Kac Moody algebras. The
lagrangian approach for such models, either fermionic or bosonic, is based in
the following idea\refto{coset,coset2}: starting from a theory with global
symmetry $G$, one constraints the Noether currents in $H$ in a gauge
invariant way.
In the case of fermionic models\refto{coset} one introduces to this end gauge
fields acting as Lagrange multipliers  which enforce the constraints. The
resulting theory is then bosonized using conventional path integral methods.

An interesting problem is posed when the subgroup $H$ is not simply
connected. In this case it is possible to classify topologically inequivalent
gauge fields by the fundamental group $\Pi_{1}(H)$. Important examples of those
groups are of the form $H={\tilde H}/Z$ where ${\tilde H}$ is a compact,
connected, simple and simply connected Lie group and $Z$ is a discrete subgroup
of the center. For such models the gauge field path integral decomposes into a
sum of terms labelled by the fundamental group of
$H$\refto{Bardakci,ellos,cm,yo}.

In this work we study a constrained fermionic model with $G=U(NK)$ and
$H=SU(N)_K/Z_{N}$ ($K$ is the Kac Moody level and $Z_{N}=Z/N$ the center of
$SU(N)$). We isolate the topologically non-trivial contribution  from the zero
charge degrees of freedom of the gauge field and decouple the latter from the
fermions. This procedure leads to the factorization of the partition function
in three terms: a partition function of fermions coupled to the non-trivial
gauge field, a partition function of a gauged Wess-Zumino-Witten (WZW) action
and a partition function of a ghost system also coupled minimally to the
background monopole. With a convenient election of the monopole background with
support at infinity (the north pole), the three subsystems mentioned above
become, separately, conformal invariant although the Virasoro generators are
affected by the background field. Using the Dotsenko-Fateev\refto{Dotfat}
analogy with a Coulomb gas, we show that the modification of the
energy-momentum tensor for each subsystem is given by
$$
\Delta T_{z z}\propto \partial_{z} j^{N-1}
$$
where $j^{N-1}$ is the current in the direction
of the ``last" generator of the Cartan subalgebra of \sun.
This result is crucial since it implies that the total change of
the energy-momentum tensor is proportional to the derivative of the
total current ${\bf J}^{N-1}$, this being zero due to the imposed
constraint. Consequently the modifications of the conformal anomalies for each
subsystem cancel each other and the total central charge of the coset system
is the same in any topological sector.
We prove this statement by computing explicitly the conformal charges of the
three subsystems for the general coset $U(NK)/SU(N)_K$.

\head{Section II}

A fermionic realization of the coset model $U(NK)/H$ is constructed
starting from a theory of $N\times K$ free Dirac fermions in which the currents
associated to the subgroup $H$ are constrained to zero \via lagrange
multipliers\refto{coset}.
The corresponding Lagrangian in $R^2$ is given by
$$
\cL={\bar \psi}_{i m}(i\slp\delta_{i j} + \lnA_{i j})\psi_{i m}
{}~~~~i,j=1,...,N;~m=1,...,K \eqno(2.1)
$$
where $\psi_{i m}$ are $N\times K$ Dirac
fermions and the gauge field $A_{\mu}$ (which plays the role of a Lagrange
multiplier) takes values in the Lie algebra of a compact connected semi-simple
Lie group $H$.
The partition function reads
$$
\cZ=\int D{\bar\psi} D\psi DA~ exp[\int d^{2}x {\bar \psi}(i\slp +
\lnA)\psi]. \eqno(2.2)
$$

In order to compactify the two dimensional manifold $R^2$ we impose appropriate
boundary conditions at infinity:
$$
\displaystyle\lim_{r\to \infty}A\to ig^{-1}dg\eqno(2.3)
$$
where
$g(\th)=\displaystyle\lim_{r\to\infty}g(r,\th)$ belongs to the loop group $LH$
of $H$.

Topologically non trivial configurations corresponds to elements $g$  not
belonging to the identity component $L_{0}H$ of $LH$. Examples of groups
supporting non trivial configurations are \refto{Cachoyyo}
$$
H={\tilde H}/Z\eqno(2.4)
$$
with ${\tilde H}$ a compact, connected, simply
connected lie group and Z a non trivial subgroup of the center. In particular
we take $H=SU(N)/Z_{N}$ ($Z_{N}=Z/N$). The fundamental group of H is
$$
\Pi_{1}(SU(N)/Z_{N})=Z_{N}\eqno(2.5)
$$
showing that $LH$ has $N$ connected
components, each of that defining a different nontrivial configuration of the
gauge field $A_{\mu}$. We will call $A_{\mu}^{(\alpha)} (~~\alpha
=0,1,...,N-1)$ a gauge field ``belonging to the topological class $\alpha$" in
the sense of equations (2.3)-(2.5).

The above discussion shows that the functional integral measure in the
partition
function (equation (2.2)) splits in the sum of N integrals, one for each
topological class $\alpha$
$$
DA~ exp(-S[A]) = \sum_{\alpha =0}^{N-1} DA^{\alpha} exp(-S[A^{\alpha}]).
\eqno(2.6)
$$

There is a problem concerning the fermionic integration in equation (2.2). This
integral can be computed using well known regularization prescriptions ($\ze$
function, heat kernel, etc.) which are valid provided the gauge field $A_{\mu}$
is topologically trivial ($\alpha =0$ in our case). Otherwise , the integral is
zero because of the zero modes of the Dirac operator. There is, however, a way
to overcome this difficulty \refto{Bardakci} which we now discuss.

A zero class (trivial) gauge field $a_{\mu}$ can be written in the form:
$$
a_{z}=iu^{-1}\partial_{z}u \eqno(2.7)
$$
$$
a_{\bar z}=iv^{-1}\partial_{\bar
z}v \eqno(2.8)
$$
where $z=x+iy$, ${\bar z}=x-iy$ and $u$ and $v$ are {\it
single-valued} fields taking values in $H$. Then we can write a general gauge
field of class $\alpha$ in the manner
$$
A_{z}^{(\alpha)}=a_{z}+u^{-1}\cA_{z}^{(\alpha)}u\eqno(2.9)
$$
$$
A_{\bar
z}^{(\alpha)}=a_{\bar z}+v^{-1}\cA_{\bar z}^{(\alpha)}v\eqno(2.10)
$$
where
$\cA_{\bar z}^{(\alpha)}$ is a {\it fixed} configuration of class $\alpha$.
For convenience we choose this field satisfying the Lorentz gauge condition
$$\partial_{\mu}\cA_{\mu}^{(\alpha)}=0.\eqno(2.11)$$ Hence the integral
$DA^{\alpha}$ reduces to an integral in the {\it topologically trivial} fields
$u$ and $v$ over the whole group manifold $H=SU(N)/Z_{N}$ (there is of course a
jacobian which we will consider later). Finally, because of the invariance of
$A_{\mu}^{(\alpha)}$ under gauge transformations in the center $Z_{N}$ we can
express the $H$ integral in terms of the SU(N) variables as follows
$$\int_{SU(N)/Z_{N}} Dg~ exp(-S[g])={1\over N}\int_{SU(N)} D\tilde g ~
exp(-S[\tilde g]).\eqno(2.12)$$

The final expression for the partition function $\cZ$ reads
$$\cZ={1\over N} \sum_{\alpha=0}^{N-1}\int_{SU(N)} D{\tilde u}D{\tilde v}
J[{\tilde u},{\tilde v}] det\left(i\slp +{\lnA}^{(\alpha)}\right)\eqno(2.13)$$
where  $A_{\mu}^{(\alpha)}$ is defined in equations (2.9),(2.10) and $J[{\tilde
u},{\tilde v}]$ is the jacobian
$$DA^{\alpha}=D{\tilde u}D{\tilde v}~ J[{\tilde u},{\tilde v}].\eqno(2.14)$$

Using standard methods it is possible to decouple the $\tilde u$ and $\tilde
v$ fields from the fermions in the determinant. In fact the following identity
holds\refto{Polyakov}
$${det\left(i\slp +{\lnA}^{(\alpha)}\right) \over det\left(i\slp +
{\lncA}^{(\alpha)}\right)}=exp(-{\cS}[{\tilde u} {\tilde
v}^{-1},{\cA}_{\mu}^{(\alpha)}]). \eqno(2.15)$$ Here
$$
{\cS}[g,\cA_{\mu}]=-KI[g]-{K\over \pi}\int d^{2}x tr\left(-i\cA_{\bar z}g^{-1}
\partial_{z}g+i\cA_{z}\partial_{\bar z}g g^{-1}-\cA_{z}g\cA_{\bar z}g^{-1}+
\cA_{z}\cA_{\bar z}\right)\eqno(2.16)$$ and $-KI[g]$ is the level $-K$
Wess-Zumino-Witten action. Of course the field $\cA_{\mu}^{(\alpha)}$ cannot be
decoupled from the fermions, since such field produces a non trivial kernel to
the Dirac operator which prevents the use of the standard methods for computing
determinants (but for a carefully treatment of the determinant of the Dirac
operator with non trivial kernel see \refto{Solo}).

Let us return to the jacobian in equation (2.14). A direct computation leads
to:
$$J[{\tilde u},{\tilde v}]=det\left(i\partial_{z} +A_{z}^{(\alpha)}\right)_
{ADJ} det\left(i\partial_{\bar z} +A_{\bar z}^{(\alpha)}\right)_{ADJ}
\eqno(2.17)$$ where the subscript {\tenit ADJ} indicate that the Dirac operator
acts in the adjoint representation of SU(N). This determinant can be related
to the determinant of the same operator in the fundamental (defining)
representation by the identity\refto{Polyakov,jacobian}(There is a similar
equation for the ${\bar z}$ component)
$${det\left(i\partial_{z} +A_{z}^{(\alpha)}\right)_{ADJ} \over det\left(
i\partial_{z} + {\cA}_{z}^{(\alpha)}\right)_{ADJ}}=  \left[{det\left(i\partial_
{z} +A_{z}^{(\alpha)}\right) \over det\left(i\partial_{z} + {\cA}_{z}^{(\alpha)
}\right)}\right]^{2C_{H}}\eqno(2.18)$$ where $C_{H}$ is the eigenvalue of the
Casimir operator in the adjoint representation (in our example $C_{SU(N)}=N$).
Using equations (2.15)-(2.18) we can write
$$J[{\tilde u},{\tilde v}]=exp(-2N{\cS}[{\tilde u}{\tilde v}^{-1},
{\cA}_{\mu}^{(\alpha)}])~det\left(i\partial_{z} + {\cA}_{z}^{(\alpha)}\right)
_{ADJ}~det\left(i\partial_{\bar z} + {\cA}_{\bar z}^{(\alpha)}\right) _{ADJ}.
\eqno(2.19)$$

Finally, after fixing the trivial gauge ${\tilde v}=1$, we can write the
partition function as $$\cZ={1\over N} \sum_{\alpha=0}^{N-1}\int_{SU(N)}~
D{\tilde g}~ det\left(i\slp + {\lncA}^{(\alpha)}\right)~
e^{(K+2N){\cS}[{\tilde g}, {\cA}_{\mu}^{(\alpha)}]}~\times$$
$$det\left(i\partial_{z} + {\cA}_{z}^{(\alpha)}\right)
_{ADJ}~det\left(i\partial_{\bar z} + {\cA}_{\bar z}^{(\alpha)}\right)
_{ADJ}.\eqno(2.20)$$ Note that in virtue of equations (2.3), (2.15) and (2.19),
the field ${\tilde g}$ has trivial boundary conditions, \ie
$$\displaystyle\lim_{r\to \infty}{\tilde g}\to I.\eqno(2.21)$$

\head{Section III}

The equation (2.20) shows that the partition function of the theory is a sum of
terms, one for each topological class. Individually each term can be
factorized in three: the partition function of $NK$ Dirac fermions coupled
to the background field
$\cA_{\mu}^{(\alpha)}$, the partition function of a gauged Wess-Zumino-Witten
theory and the partition function of $N^2-1$ pair of Fadeev-Popov ghosts also
coupled to $\cA_{\mu}^{(\alpha)}$.
As we commented before, the non-trivial topology has a deep impact in the Dirac
operator: its kernel becomes non-trivial. In order to study this problem more
carefully let us find the square-integrable solutions of the differential
equation
$$\left(i\slp + {\lncA}^{(\alpha)}\right)\psi=0~,~~~~~\psi=\left[\matrix{
\psi_{1}\cr\psi_{2}\cr\vdots\cr\psi_{N}\cr}\right]\eqno(3.1)$$
(we are considering for simplicity the case with one flavor, the generalization
to the case of $K$ flavors is straightforward).
First at all we have to chose a representative background field. In the
preceeding section we mapped the problem initially defined on $SU(N)/Z_{N}$
over its universal covering $SU(N)$. In this manifold the different components
of the loop group $L(SU(N)/Z_{N})$ are represented by multivalued group-valued
functions
$$g~:~[0,2\pi]\to SU(N)~~~~/~g(0)=h\cdot g(2\pi)\eqno(3.2)$$
where $h$ is an element of the center $SU(N)$ (\ie $h^N=1$). Its
easy to verify that the more simpler configurations satisfying (3.2) are
$$g_{n}(\theta)=exp(i{n\over N}T^{N-1}\theta)\eqno(3.3)$$
where
$$T^{N-1}=\pmatrix{1&0&0&\ldots&0\cr
                   0&1&0&\ldots&0\cr
                   0&0&1&\ldots&0\cr
                   \vdots&\vdots&\vdots&\ddots&\vdots\cr
                   0&0&0&\ldots&1-N\cr}\eqno(3.4)$$
is the $N-1$ Cartan generator of $SU(N)$ and $n\equiv mod(N)$ label the
topological sector. Finally our election of the background gauge field
is\refto{letogus}
$$\eqalign{\cA_{\mu}^{(n)}&=-ia(\rho) g_n^{-1}\partial_{\mu}g_n\cr
                          &={n\over N}a(\rho)\partial_{\mu}\theta T^{N-1}}
\eqno(3.5)$$
where $a(\rho)$ is a function which vanishes at the origin and has the
asymptotic behaviour
$$\displaystyle\lim_{\rho\to\infty}a(\rho)\to 1\eqno(3.6)$$
enforced by the boundary condition equation (2.3). One can verify that this
election satisfies the Lorentz gauge condition equation (2.11) and consequently
it can be written as a curl
$$\cA_{\mu}^{(n)}=-\epsilon_{\mu \nu}\partial_{\nu}K(\rho)T^{N-1}.\eqno(3.7)$$
Here $K(\rho)$ is defined as
$$K(\rho)={n\over N}\int_0^{\rho} {a(r)\over r}dr\eqno(3.8)$$
and has the following behaviour
$$\displaystyle\lim_{\rho\to\infty}K(\rho)\to \rho^{n\over N}.\eqno(3.9)$$

Then, in terms of the chiral components $\psi_{L}$ and $\psi_{R}$, the equation
(3.1) takes the form
$$\partial_{z}\left( e^{-K(\rho)T^{N-1}}\psi_{R}\right)=0\eqno(3.10)$$
$$\partial_{\bar z}\left( e^{K(\rho)T^{N-1}}\psi_{L}\right)=0\eqno(3.11)$$
with solutions
$$\psi_{R}(x,y)=e^{K(\rho)T^{N-1}}\chi_{R}(\bar z)\eqno(3.12)$$
$$\psi_{L}(x,y)=e^{-K(\rho)T^{N-1}}\chi_{L}(z)\eqno(3.13)$$
where $\chi_{R}$ ($\chi_{L}$) is an arbitrary anti-holomorphic (holomorphic)
N-components spinor. However by requiring square-integrability the possible
values of $\chi_{R}$ and $\chi_{L}$ have severe constrains.

There is a subtlety concerning this issue. Because we are dealing with the
compactified plane, the inner product of spinors is different from the usual
one. In particular the inner product induced by the standard metric of the
sphere is\refto{nielsen}
$$(\psi,\chi)=\int {2\over (1+\vec x^2)}d^2\vec
x~\psi^{\dagger}(\vec x)\chi(\vec x).\eqno(3.14)$$

Writing the $n$ as
$$n=r+cN~,~~r,c\in Z~,~~0\leq r<N\eqno(3.15)$$
($r$ is the ``topological charge") and using the equations (3.12), (3.13) and
(3.14) we find a set of basis vectors for the kernel of $\slD$ (we consider
$c>0$)

Case (I) $r\not=0$
$$\psi_{L}^{(q)}=e^{(1-N)K(r)}\left[\matrix{
0\cr\vdots\cr 0\cr z^{q}\cr}\right]~,~~~~q=0,1,\dots,n-c-1
\eqno(3.16)$$
$$\psi_{R}^{(i,p)}=e^{-K(r)}\left[\matrix{
0\cr\vdots\cr {\bar z}^{p}\cr\vdots\cr 0\cr}\right]~,~~~~\cases
{i=1,2,\dots,N-1~~ {\rm (labels~the~rows)}\cr p=0,1,\dots,c-1,c.\cr}
\eqno(3.17)$$

Case (II) $r=0$
$$\psi_{L}^{(q)}=e^{(1-N)K(r)}\left[\matrix{
0\cr\vdots\cr 0\cr z^{q}\cr}\right]~,~~~~q=0,1,\dots,n-c-1
\eqno(3.17)$$
$$\psi_{R}^{(i,p)}=e^{-K(r)}\left[\matrix{
0\cr\vdots\cr{\bar z}^{p}\cr\vdots\cr 0\cr}\right]~,~~~~\cases
{i=1,2,\dots,N-1~~{\rm (labels~the~rows)}\cr p=0,1,\dots,c-1.\cr}
\eqno(3.18)$$

Finally we can compute the index of the Dirac operator and we find
$$index~\slD=dim~ker~D_{z}-dim~ker~D_{\bar z}=\cases {(N-1)-r,&if $r\not=0$
\cr 0,&if $r=0$\cr}\eqno(3.19)$$
(for $K$ flavors this value is multiplied by $K$).

Note that even when the number of zero modes is $n$-dependent, the index {\it
only depends on} $n\equiv mod(N)$, the topological charge.

There are some direct conclusions we can obtain from these results. As we
mentioned before, the existence of zero modes of the Dirac operator enforces
constraints on the fermionic correlation functions. These features can be
explained easily in the path-integral formalism. The fermionic path-integral
can be performed by expanding the fermions in a base of eigenvectors
of the Dirac operator with Grassmann variables as coefficients, and finally,
integrating the Grassmann variables with the Berezin rules. However those
Grassmann coefficients associated to the zero modes will be absents in the
fermionic action. Therefore the only non-vanishing correlation functions are
those which can provide the necessary number of Grassmann variables to make the
integrals non-zero. In general, if we define the following family of
``chirality
changing" fermionic operators
$$\alpha_{R}^{i j}(z)={\bar \psi}^{i}_{R}(z) \psi^{j}_{R}(z)\eqno(3.20)$$
$$\alpha_{L}^{i j}(z)={\bar \psi}^{i}_{L}(z) \psi^{j}_{L}(z)\eqno(3.21)$$
we can derive easily the following results. For $r\not=0$ the non-zero
correlation  functions of $\alpha$ operators contains a number of
$\alpha_{R}^{i j}(z)$ bilinears
which exceeds in $N-1-r$ the number of $\alpha_{L}^{i j}(z)$ bilinears.
Moreover
the minimum number of $\alpha_{L}^{i j}(z)$ operators presents must be $n-c$.
For $r=0$ the non-vanishing correlation functions contains the same number of
$\alpha_{R}^{i j}(z)$ bilinears than $\alpha_{L}^{i j}(z)$ bilinears, with a
minimum of $n-n/N$ of each class.

Foe the special case of $N=2$ there are only one non-trivial topological sector
which we can parametrize with $n=1$. For this case there are two fermionic zero
modes, one of left chirality and the other of the opposite chirality. The
non-vanishing correlation functions will contain the same number of
$\alpha_{R}^{i j}(z)$ bilinears than the $\alpha_{L}^{i j}(z)$ bilinears.

\head{Section IV}
The aim of this section is to compute the conformal anomaly of this ``twisted"
model. We will study the three factors of the partition function separately
(\ie the bosonic, the fermionic and the ghost sector). Of course each of this
sectors are not independently conformal invariant for an arbitrary background
gauge field. However following \Ref{yo,cm} we can overcome this difficulty by
choosing the {\it arbitrary} background field $\cA_{\mu}$ concentrated at the
infinity (with support at the infinity). With this election the three sectors
above mentioned becomes {\it separately} conformal invariant on $R^2 = S_2 -
\{north~pole\}$. Moreover we will show that the effect of the background field
in the three cases is a modification of the conformal properties as in the
Dotsenko-Fateev's
model\refto{Dotfat}. We will also see that due to the constraints of the
theory the shifts of the conformal anomalies of the three sectors add to zero,
and the total conformal anomaly of the model is independent of the topological
sectors\refto{yo}.

\vskip 1.0cm
\noindent{\bf a) The Case $H=SU(2)/Z_2$}

Let us first consider the easier case $SU(2)/Z_2$. Once solved this problem
the generalization to the case $SU(N)/Z_N$ is straightforward.

We begin with the analysis of the bosonic action. The bosonic effective
action is given by
$${\cS}[g,\cA_{\mu}]=kI[g]-{k\over \pi}\int d^{2}x tr\left(-i\cA_{\bar z}g^{-1}
\partial_{z}g+i\cA_{z}\partial_{\bar z}g g^{-1}-\cA_{z}g\cA_{\bar z}g^{-1}+
\cA_{z}\cA_{\bar z}\right)\eqno(4.1 a)$$
where $kI[g]$ is the level $k$ Wess-Zumino-Witten action ($k=-(K+4)$ in our
case) and the background field takes the form
$$\cA_{\mu}^{(n)}=-\epsilon_{\mu \nu}\partial_{\nu}K(\rho)\sigma^3.\eqno(4.2 a)
$$
Although the presence of this field breaks the chiral $SU(2)_{L}\times
SU(2)_{R}$ invariance, the symmetries in the Cartan subgroup
$U(1)_{L} \times U(1)_{R}$ remains unbroken. The currents associated with this
symmetry are
$$J_{z}^3=j_{z}^{3} + ik~tr(g^{-1}\cA_{z}
g\sigma^3 + \cA_{z}\sigma^3)\eqno(4.3 a)$$
$$J_{\bar z}^3=j_{\bar z}^{3} + ik~tr
(g\cA_{\bar z}g^{-1}\sigma^3 + \cA_{\bar z}\sigma^3)\eqno(4.4 a)$$
where
$$j_{z}^{3}=-k~tr(g^{-1}\partial_{z}g \sigma^3)\eqno(4.5 a)$$
$$j_{\bar z}^{3}=-k~tr(g\partial_{\bar z}g^{-1} \sigma^3)\eqno(4.6 a)$$
and we use the gauge condition equation (2.11). Parallel to the analysis of
the abelian case (see Dotsenko-Fateev\refto{Dotfat} and \Ref{yo}), we easily
verify that the second term of each current measures the background charge (in
the Cartan algebra direction) created at infinity by the gauge field
$\cA_{\mu}$.
In fact, if the operators $V_{q_{i}}(z_i)$ creates
respectively a $\sigma^3$ charge $q_{i}$ at the point $z_{i}$ ($i=1,...,n$),
the {\it total} $\sigma^3$ charge of the state
$$|\{q_i,z_i\}>=\prod_{i=1}^{n}V_{q_{i}}(z_{i})|0>\eqno(4.7 a)$$
is computed by
$$\eqalign{Q^3_{z}&={1\over 4\pi i}\oint dz~J_{z}^{3}|\{q_i,z_i\}>\cr &={1\over
4\pi i}\oint dz~j_{z}^{3}|\{q_i,z_i\}> +{1\over
2\pi}k\oint\cA_{\mu}dx^{\mu}|\{q_i,z_i\}> \cr &=\left(\sum_{i=1}^{n} q_{i} +
{k\over 2}\right)|\{q_i,z_i\}>.\cr}\eqno(4.8 a)$$
Hence, the neutrality of the charge enforces the following constraint on the
vertex operators
$$\sum_{i=1}^{n} q_{i} = - {k\over 2} = \left({K\over 2}+2\right) ~~~for~
k=-(K+4).\eqno(4.9 a)$$
This relation (equation $(4.9 a)$) changes the conformal dimensions of the
vertex operators and consequently the energy-momentum tensor differs from  the
one of a pure WZW theory. The analogy of this model with the Dotsenko and
Fateev's Coulomb gas suggests that such a modification is given by the addition
of a term:
$$\Delta T_{z z}\propto \partial_{z} j^{3}_{z}.\eqno(4.10 a)$$

Let us now recall the free field
representation of the WZW theory\refto{wakimoto}. The kernel of this
method is to find a realization of the affine algebra by means of free fields.
For the SU(2) group we can write the affine currents in terms of three free
fields, $\phi$, $\mu$ and $\nu$, and they reads
$$j_{z}^{3}=i\sqrt{2(k+2)}\partial_{z}\phi + 2\partial_{z}\mu\eqno(4.11 a)$$
$$j_{z}^{+}=\sqrt{2}\partial_{z}\nu~ e^{-2(\mu - i\nu)}\eqno(4.12 a)$$
$$j_{z}^{-}=\left(-i\sqrt{2}(k+2)\partial_{z}\mu+2\sqrt{k+2}\partial_{z}\phi-
\sqrt{2}(k+1)\partial_{z}\nu\right) e^{-2(\mu - i\nu)}.\eqno(4.13 a)$$
In order to reproduce the conformal anomaly of the WZW model the fields $\phi$,
 $\mu$ and $\nu$ are coupled to the Gaussian curvature. Their actions can be
written as
$$\cS_{\Phi}={1\over 2\pi}\int
d^{2}x~\left\{(\partial\Phi)^{2}+Q_{\Phi}R\Phi\right\}~~~~~~
\Phi=\phi, \mu, \nu\eqno(4.14 a)$$
where $R$ is the scalar curvature and the ``charges" $Q_{\Phi}$ takes the
values
$$Q_{\phi}=-{i\sqrt{2}\over
\sqrt{k+2}}~,~~~~Q_{\mu}=-1~,~~~~Q_{\nu}=i.\eqno(4.15 a)$$
The corresponding energy-momentum tensors and conformal charges are
$$T_{\Phi}=-2(\partial_{z}\Phi)^{2}+Q_{\Phi}\partial_{z}^{2}\Phi\eqno(4.16 a)$$
$$c_{\Phi}=1+3Q_{\Phi}^2.\eqno(4.17 a)$$
Returning to our model, a WZW model with a minimal coupling to a background
field located at infinity, it is now easy to describe the effect of the
background field in terms of the free field realization. In fact, we can
show\refto{Dotfat, yo} that, instead of working with the currents $J_{z}^{3}$
and $J_{\bar z}^{3}$  (equations $(4.3 a)$ and $(4.4 a)$) we can use the free
currents $j_{z}^{3}$ and $j_{\bar z}^{3}$ but defining a ``new" vacuum state
$|k>$ as
$$
|k>=e^{-i\sqrt{2(k+2)}\phi(\infty)} e^{-2\mu(\infty)}|0>.\eqno(4.18 a)
$$
This vertex insertion creates a charge $k$ at the infinity and therefore
reproduces the contribution of the background field in equation $(4.8 a)$.
Moreover, we can include this insertion in the actions $(4.14 a)$ by a simple
redefinition of the charges $Q_{\Phi}$ which takes the form
$$
Q_{\phi}={ik\over
\sqrt{2(k+2)}}~,~~~~Q_{\mu}=0~,~~~~Q_{\nu}=i.\eqno(4.19 a)
$$
Note that this procedure is equivalent to add to the energy-momentum tensor the
term
$$
\Delta T(z)_B={1\over 2}\partial_{z}j^3_z(z)\eqno(4.20 a)
$$
in accordance with equation $(4.10 a)$.
Finally using equation $(4.17 a)$ we find the conformal charge of this model
$$c_{B}=-{3k^2\over 2(k+2)}={3(N+4)^2\over 2(N+2)}~~~for~k=-(N+4).\eqno(4.21 a)
$$

We can also study the fermionic action in a close related way. With the
election $(4.2 a)$ of the gauge field (and after a harmless rescaling of the
fermions) the fermionic action is the sum of $K$ terms of the form
$$S_{F}={1\over 4\pi}\int d^{2}x {\bar\psi}^{(1)}\left(i\slp+\lnA\right)
\psi^{(1)} + {1\over 4\pi}\int d^{2}x {\bar\psi}^{(2)}\left(i\slp-\lnA
\right) \psi^{(2)}\eqno(4.22 a)$$
where
$$A_{\mu}=-\epsilon_{\mu \nu}\partial_{\nu}K(\rho)\eqno(4.23 a)$$
and the two SU(2) color components decouple. We can then analyze each color
separately. (We will analyze the action corresponding to $\psi^{(1)}$, all
the results are valid for $\psi^{(2)}$ replacing $A_{\mu}\to -A_{\mu}$)
For the first action we have a conservation equation (vectorial symmetry) and
an  anomalous equation (chiral symmetry)
$$\partial_{\mu} j_{\mu}=0~,~~~~~j_{\mu}={\bar\psi}\gamma_{\mu}\psi\eqno
(4.24 a)$$
$$\partial_{\mu} j_{\mu}^{5}=-2\epsilon_{\mu \nu}F_{\mu \nu}~,~~~~~
j_{\mu}^{5}={\bar\psi}\gamma_{\mu}\gamma^{5}\psi.\eqno(4.25 a)$$
However, because the gauge field satisfies the Lorentz gauge condition, we can
construct a new pair of conserved currents, one holomorphic and the other
anti-holomorphic
$$\partial_{\bar z} J_{z}=0~,~~~~J_{z}=i{\bar
\psi}_{R}\psi_{L}+4A_{z}\eqno(4.26 a)$$
$$\partial_{z} J_{\bar z}=0~,~~~~J_{\bar z}=i{\bar
\psi}_{R}\psi_{L}-4A_{\bar z}.\eqno(4.27 a)$$
Note the similarity of this currents with the bosonic currents $J_{z}^{3}$ and
$J_{\bar z}^{3}$ (equations $(4.3 a)$ and $(4.4 a)$). In fact we can proceed in
the same way we derive the equations $(4.7 a)$-$(4.10 a)$ and show that the
effect of the ``anomalous" term in the currents $J_{z}$ and $J_{\bar z}$ is to
create a Dotsenko and Fateev's charge ${i\over 2}$ at infinity\refto{yo}.
As in the bosonic case we can incorporate this effect in the energy-momentum
tensor by adding a term
$$
\Delta T_{F}\propto \pz j_{z~F}\eqno(4.28 a)
$$
where $j_{z~F}=i{\bar \psi}_R\psi_L$ is the left
``number" current. Finally using the rules of abelian
bosonization\refto{FMS,cuerdistas} we can compute the changes in the
Virasoro central charge employing again the equation $(4.17 a)$ with $Q={i\over
2}$. Finally adding the contribution of the fermion  $\psi^{(2)}$ (replacing
$Q\to -Q$) and summing over the $K$ flavors  we find that the conformal charge
for the fermionic action is
$$
c_{F}=K\left({1\over 4} + {1\over 4}\right)={K\over 2}.\eqno(4.29 a)$$

The former argument can be easily extended to study the the contribution to the
conformal anomaly of the ghost action in equation (2.20).  The determinant in
the adjoint representation can be written as the path-integral
$$det\left(i\partial_{z} + {\cA}_{z}^{(\alpha)}\right) _{ADJ}=\int D\bar\xi
D\xi e^{-{1\over 2\pi}\int tr(\bar\xi \cD_{z} \xi)}\eqno(4.30 a)$$
where $\bar\xi~,~\xi$ are a pair ghost-antighost taking values in the Lie
algebra of $SU(2)$ and $\cD_{z}$ is defined by
$$\cD_{z}\xi=i\partial_{z}\xi + [A_{z},\xi].\eqno(4.31 a)$$
We can expand the ghosts fields in the Cartan-Weyl basis $(\sigma^+, \sigma^-,
\sigma^3)$ introducing the three pairs of ghosts $(b^{+},c^{+})~,~(b^{-},
c^{-})$ and $(b^{3},c^{3})$
$$\bar\xi=b^{+}\sigma^{-}+b^{-}\sigma^{+}+b^{3}{\sigma^{3} \over
\sqrt{2}}\eqno(4.32 a)$$
$$\xi=c^{+}\sigma^{-}+c^{-}\sigma^{+}+c^{3}{\sigma^{3} \over
\sqrt{2}}\eqno(4.33 a)$$
(the generators are normalized to
$tr~(\sigma^{+}\sigma^{-})=tr~({\sigma^3)}^2=1$ and any other combination equal
to zero).

In this representation the Lagrangian in $(4.30 a)$ is diagonal in the ghost
fields and takes the form
$$S[{\vec b},{\vec c}]={1\over 2\pi}\int d^{2}x\left\{
b^{+}_{z}(i\partial_{\bar z}+2A_{\bar z})c^{-}_{z}+b^{-}_{z}(i\partial_{\bar
z}-2A_ {\bar z})c^{+}_{z}+b^{3}_{z}(i\partial_{\bar
z})c^{3}_{z}\right\}.\eqno(4.34 a)$$
(Of course there is an identical expression for the $det\left(i\partial_z
 + {\cA}_z^{(\alpha)}\right)_{ADJ}$). The fields $A_{z}$ and $A_{\bar
z}$ are the same defined in equation $(4.23 a)$ and
$(b^{i}_{z},~c^{i}_{z})~(i=+,-,3)$ are ghosts of conformal dimensions 1 and 0
respectively. There is a fundamental difference between this system and the
fermionic one (equation $(4.22 a)$). Owing to the tensorial character of the
ghosts they have a non-trivial coupling with the metric. In fact although the
gauge current in both systems are conserved the chiral current for the ghosts
suffers two types of anomalies, the usual chiral anomaly  and a gravitational
anomaly proportional to the Gaussian curvature of the
two-manifold\refto{Fujikawa,FMS,cuerdistas}. For example for the system
$(b^{+},~c^{-})$ we have
$$\partial_{\mu}
j_{\mu}=0~,~~~~{\vec j}=\left(-b_{z}^{+}c_{z}^{-}-b_{\bar z}^{+} c_{\bar
z}^{-}~,~-ib_{z}^{+} c_{z}^{-}-ib_{\bar z}^{+}c_{\bar
z}^{-}\right)\eqno(4.35 a)$$
$$\partial_{\mu} j_{\mu}^{5}=-4\epsilon_{\mu \nu}F_{\mu
\nu}-2\sqrt{g}R~,~~~~{\vec j}^{5}=\left(-b_{z}^{+}c_{z}^{-}+b_{\bar z}^{+}
c_{\bar z}^{-}~,~-ib_{z}^{+} c_{z}^{-}+ib_{\bar z}^{+}c_{\bar
z}^{-}\right).\eqno(4.36 a)$$
In a conformally flat metric $g_{\mu \nu}=e^{4\eta}\delta_{\mu \nu}$ we can
also write a new pair of conserved currents as in equations $(4.26 a)$ and
$(4.27 a)$
$$\partial_{\bar z}
J_{z}=0~,~~~~J_{z}=ib_{z}^{+}c_{z}^{-}+8A_{z}+4i\partial_{z}\eta\eqno(4.37 a)$$
$$
\partial_{z} J_{\bar z}=0~,~~~~J_{\bar z}=ib_{\bar z}^{+}c_{\bar z}^{-}-
8A_{z}-4i\partial_{\bar z}\eta\eqno(4.38 a)
$$
With the help of the Riemann-Roch theorem we can prove that the anomalous terms
in the currents alter the balance of the ghosts charge by creating a
background charge\refto{yo} $Q_1=i+i$ (from the gauge field and from the
Gaussian curvature respectively).  The analysis of the system $(b^-,c^+)$ is
similar. In virtue of the opposite sign of the gauge field in equation
$(4.34 a)$
the background charge in this case is $Q_2=i-i$.  Finally for the free ghost
system $(b^3,c^3)$ we have $Q^3=0+i$.
Once again we can realize these effects by adding to the energy-momentum tensor
a term
$$
\Delta T_G\propto\pz j^3_{z~G}\eqno(4.39 a)
$$
where $j^3_{z~G}=i(~2b^+_zc^-_z -2b^-_zc^+_z + b^3_zc^3_z~)$.
Using  the equation $(4.17 a)$ with the above mentioned
values of the charges we obtain the total ghost central charge
$$
c_{G}=-12.\eqno(4.40 a)
$$

The total conformal charge of this coset model is the sum of the values
$(4.21 a)$, $(4.29 a)$ and $(4.40 a)$ and the answer is
$$
c={2(K-1)^2\over K+2}+1\eqno(4.41 a)
$$
which corresponds exactly to the usual value of the central charge of the coset
model  $U(2K)/SU(2)_K \sim SU(K)_2\times U(1)$. That is we find that the
conformal anomaly is independent of the topological background charge.

At the light of the above discussion this result became expected. In fact, we
have proved that for each subsystem \ie bosonic, fermionic or ghost, the effect
of the monopole is a shift of the energy-momentum tensor in a quantity
$$
\Delta T_{B,F,G}\propto \partial_{z} j^{3}_{B,F,G}\eqno(4.42 a)
$$
where $B, F, G$ stands for bosons, fermions and ghost respectively. Thus the
change of the total energy-momentum tensor of the theory is proportional to the
derivative of the \total~ current in the direction $\sigma^3$. But this is one
of the currents we are constraining to vanish with the Lagrange multiplier \ie
$j^3_{total} \equiv 0$. Hence once proved that the modification of the
energy-momentum due to the presence of the monopole is of the form $(4.42 a)$,
the result $(4.41 a)$ is natural. Note that this result, the independence of
the conformal anomaly on the topological sector, is a property of the whole
theory but each subsystem separately suffers a change. In particular the
primary fields of each subsystem have topology-dependent conformal dimensions.
Moreover the vacuum expectation values of the {\it original} fermionic fields
depends on the topological charge as we mentioned in the previous section.

In the next subsection we will show that these results are also valid for
the general coset $U(NK)/SU(N)_K  \sim SU(K)_N\times U(1)$.

\vskip 1.0cm
\noindent{\bf b) The Case $H=SU(N)/Z_N$}

Now we can generalize the former results to the case $H=SU(N)/Z_N$. In this
case there are $N-1$ different topological sectors and not just one as in the
$SU(2)/Z_2$.
We begin with the study of the gauged WZW action (2.16).
First we need to recall, as in the earlier case, the free field realization of
the affine Kac-Moody algebra associated to \sun\refto{wakimoto}.

Let $\{ h_i,~e_i,~f_i;~~i=1,\cdots,N-1\}$ the Chevalley basis of \sun~ defined
by the relations
$$
\eqalign{[e_i,f_j]&=\delta_{i j} h_i\cr
         [h_i,e_j]&=a_{i j} e_j\cr
         [h_i,f_j]&=-a_{i j} f_j\cr}\eqno(4.1 b)
$$
where
$$
a_{i j}={2(\alpha_i\cdot \alpha_j)\over (\alpha_j\cdot \alpha_j)}\eqno(4.2 b)
$$
is the Cartan matrix. Our convention for the simple roots $\alpha_j$
is\refto{georgi}
$$
\alpha_j=-\sqrt{{j-1\over 2j}}~\ee_{j-1}+\sqrt{{j+1\over 2j}}~\ee_{j}\eqno(4.3
b)
$$
where $\{\ee_1,\cdots,\ee_{N-1}\}$ is the canonical base of $\cR^{N-1}$. The
simple roots are normalized to
$$
2(\alpha_i\cdot \alpha_j)=2\delta_{i j}-\delta_{i+1 j}-\delta_{i
j+1}.\eqno(4.4 b)
$$
Then the affine Kac-Moody algebra associated to \sun~ is defined by the
following OPE
$$
\eqalign{
H_i(z)H_j(\omega)&=2(\alpha_i\cdot\alpha_j){k\over (z-\omega)^2}+\cdots\cr
H_i(z)E_j(\omega)&={1\over z-\omega}a_{i j}E_j(\omega)+\cdots\cr
H_i(z)F_j(\omega)&=-{1\over z-\omega}a_{i j}F_j(\omega)+\cdots\cr
E_i(z)F_j(\omega)&={k\over (z-\omega)^2}\delta_{i j} + {1\over z-\omega}
\delta_{i j}H_j(\omega)+\cdots\cr}\eqno(4.5 b)
$$
where $k$ is the Kac-Moody level.

For each positive root $\alpha_{p p+q}$ ($\alpha_{p p+q}=\alpha_p+\alpha_{p+1}+
\cdots+ \alpha_{p+q-1};~~p+q\leq N$) we introduce\refto{wakimoto}  a pair of
scalar bosons $\mu_{\alpha},~ \nu_{\alpha}$ whose dynamics is governed by the
Liouville action
$$
\cS_{\alpha}={1\over 2\pi}\int\left\{(\partial \mu_{\alpha})^2 -
R\mu_{\alpha}\right\}d^2x + {1\over 2\pi}\int\left\{
(\partial \nu_{\alpha})^2 +iR\nu_{\alpha}\right\}d^2x.\eqno(4.6 b)
$$
The OPE of these fields is given by
$$
\mu_{\alpha}(z)\mu_{\beta}(\omega)=\nu_{\alpha}(z)\nu_{\beta}(\omega)=
-{1\over 4}\delta_{\alpha \beta} \ln{(z-\omega)}.\eqno(4.7 b)
$$

We also introduce\refto{wakimoto} another set of scalar bosons
$\phi_i,~i=1,\cdots,N-1$, one  for each Cartan generator, whose action is also
a Liouville action
$$
\cS_{\vec \phi}={1\over 2\pi}\int d^2x\left\{\partial {\vec \phi}\cdot\partial
{\vec \phi} - i{2\sqrt{2}\over \sqrt{K+N}} R \rh\cdot{\vec \phi}\right\}
\eqno(4.8 b)
$$
where $\rh = {1\over 2}\sum_{\alpha>0} \alpha$. The OPE of these fields is
given by
$$
\phi_i(z)\phi_j(\omega)=-{1\over 4}\delta_{i j}\ln{(z-\omega)}.\eqno(4.9 b)
$$

Hence we can realize the entire affine current algebra in terms of the fields
$\mu_{\alpha},~\nu_{\alpha},~\phi_i$. For example the currents in the Cartan
subalgebra (luckily these are the only currents we need) can be written as
$$
H_i(z)=\sum_{\alpha>0}4(\alpha_i\cdot\alpha)\pz\mu_{\alpha}+i2\sqrt{2(K+N)}
\alpha_i\cdot{\vec \phi}.\eqno(4.10 b)
$$
The energy momentum tensor of this theory is given by
$$
\eqalign{
T(z)&=\sum_{\alpha>0}T_{\alpha}(z)+T_{\vec \phi}(z)\cr
    &=\sum_{\alpha>0}\left\{-2(\pz\mu_{\alpha})^2-\pz^2\mu_{\alpha}-
2(\pz\nu_{\alpha})^2+i\pz^2\nu_{\alpha}\right\}+\cr
    &\phantom{\sum_{\alpha>0}}+\left\{-2\pz{\vec \phi}\cdot\pz{\vec \phi}-
i{2\sqrt{2}\over \sqrt{K+N}}\rh \cdot{\vec \phi}\right\}\cr}\eqno(4.11 b)
$$
which satisfies a Virasoro algebra with a central charge
$$
c={k(N^2-1)\over k+N}.\eqno(4.12 b)
$$
Is important to mention that this energy-momentum tensor, computed by varying
the actions $(4.6 b)$, $(4.8 b)$ respect of the metric is the same one obtain
in the Sugawara construction with the currents $H_i,~E_{\alpha},~F_{\alpha}$.
Hence we conclude that because this model has the same current algebra and the
same Virasoro algebra that the $SU(N)_k$ WZW theory, both theories are
equivalents.

Now we can return to our problem: How the presence of the topological
background field $\cA_{\mu}$ in the action (2.16) affects the conformal
anomaly. The answer is not difficult. By the same arguments we use for the case
{\it N}=2 is easy to see that the effect of the background field is the
appearance in the energy-momentum tensor of an additional term
$$
\Delta T(z)={n\over N}\pz \jz\eqno(4.13 b)
$$
where $\jz = tr\left(k g^{-1}\pz g~T^{N-1}\right)$
is the Kac-Moody current in the direction $T^{N-1}$, $k=-(M+2N)$ is the level
and $n=1,\cdots,N-1$ is the topological charge.
We can express this quantity in terms of the free fields $\mu_{\alpha},
{}~\nu_{\alpha},~\phi_i$ and add the result to the free field representation of
the energy-momentum tensor equation $(4.11 b)$. In fact noting that
$$
T^{N-1}=\sum_{m=1}^{N-1} mh_m\eqno(4.14 b)
$$
we can write the current $\jz$ as
$$
\jz(z)={1\over 2}\sum_{j=1}^{N-1} jH_j(z).\eqno(4.15 b)
$$
And hence using the representation $(4.3 b)$ for the roots we find
$$
\jz(z)=N\sum_{j=1}^{N-1}\pz \mu_{i N} + i2\sqrt{(k+N)N(N-1)}\pz \phi_{N-1}.
\eqno(4.16 b)
$$
Finally the energy-momentum tensor of the ``twisted" theory takes the form
$$
\eqalign{
T_{B}^{(n)}=&\sum_{i<j<N}\left\{-2(\pz\mu_{i j})^2-\pz^2\mu_{i j} -
2(\pz\nu_{i j})^2+i\pz^2\nu_{i j}\right\}+\cr
           +&\sum_{i=1}^{N-1}\left\{-2(\pz\mu_{i N})^2-(1-n)\pz^2\mu_{i N} -
2(\pz\nu_{i N})^2+i\pz^2\nu_{i N}\right\}+\cr
           -&2(\pz {\vec \phi}\cdot\pz {\vec \phi}) - i{2\sqrt{2}\over
\sqrt{k+N}} \left(\rh-n(K+N)\sqrt{{N-1\over 2}}\ee_{N-1}\right)\cdot\pz^2 {\vec
\phi}^{N-1} .\cr}\eqno(4.17 b)
$$
The conformally anomaly can be easily computed using equation $(4.17 a)$ and we
obtain the result (for $k=-(K+2N)$)
$$
c_{B}^{(n)}={(K+2N)(N^2-1)\over K+N} + 3n^2{(K+2N)(N-1)\over N}.\eqno(4.18 b)
$$

Now let us analyze the fermionic system. The action can be written as
$$
\cS_F=\sum_{j=1}^{N-1}\sum_{m=1}^M {1\over 4\pi}\int d^2
{\bar \psi}_m^j(i\slp + \lnA)\psi_m^j +
     \sum_{m=1}^M {1\over 4\pi}\int d^2 {\bar \psi}_m^N(i\slp +(1-N)\lnA)
\psi_m^N\eqno(4.19 b)
$$
where $A_{\mu}$ is a gauge field with a {\it fractional} topological charge
$q=n/N$. As in the case {\it N}=2 this model reduces to the case of {\it KN}
abelian  fermionic system in presence of a fractionally charged topological
field. Taking into account the chiral anomaly and using the Lorentz gauge
condition (2.11) we can write $N$ pairs of conserved currents, one for each
color
$$
\partial_{\bar z} j_{z~F}^j=0~~~~j=1,\cdots,N\eqno(4.20 b)
$$
$$
\partial_{z} j_{\bar z~F}^j=0~~~~j=1,\cdots,N\eqno(4.21 b)
$$
where
$$
\eqalign{
j_{z~F}^j&=i{\bar \psi}_R^j\psi_L^j~~~~j=1,\cdots,N-1\cr
j_{z~F}^N&=i(1-N){\bar \psi}_R^N\psi_L^N\cr
j_{\bar z~F}^j&=i{\bar \psi}_L^j\psi_R^j~~~~j=1,\cdots,N-1\cr
j_{\bar z~F}^N&=i(1-N){\bar \psi}_L^N\psi_R^N.\cr}\eqno(4.22 b)
$$
Hence  we can compute the conformal anomaly using again formula $(4.17 a)$ with
the values of the charges $Q_m^i=q,~i<n$ and $Q_m^N=q(1-N)~~(m=1,\cdots,M)$.
The result is
$$
c_F^{(n)}=KN-3n^2{K(N-1)\over N}.\eqno(4.23 b)
$$

Finally we study the ghost action which appears when we exponentiate the
determinant of the Dirac operator in the adjoint representation using Grassmann
variables. To do this we introduce for each root $\alpha$ and for each Cartan
generator $h^i$ of \sun~ a ghost system, $(b^\alpha,~c^\alpha)$ and
$(b^i,~c^i)$
respectively, of conformal dimensions (1,0).  Using the commutation relations
$$
\eqalign{
[T^{N-1},E_{i j}]&=\cases{0  &$i,j<N$\cr
                         -N  &$i<j=N$\cr
                          N  &$j<i=N$\cr}\cr
    [T^{N-1},h_i]&=0~~~i=1,\cdots,N-1\cr}\eqno(4.24 b)
$$
we can write the ghost action as
$$
\eqalign{
\cS_{Gh}=&\sum_{i=1}^{N-1}b_{Ni}(i\pz+NA_z)c_{Ni}+\sum_{i=1}^{N-1}
               b_{iN}(i\pz-NA_z)c_{iN}+\cr
         +&\sum_{p=1}^{(N-1)^2}b_p~i\pz c_p + h.c.\cr}\eqno(4.25 b)
$$
Once more is easy to follow the steps of the case {\it N}=2 to compute the
conformal anomaly. We use again  the formula $(4.17 a)$ with the values of the
background charges $Q_{jN}=i(1+n)$, $Q_{Nj}=i(1-n)$ and $Q_p=i$ for the ghost
systems ($b_{Nj},~c_{Nj}$), ($b_{jN},~c_{jN}$) and ($b_{p},~c_{p}$)
respectively ($j=1,\cdots,N=1;~p=1, \cdots,(N-1)^2$) and we obtain the result
$$
c_{G}^{(n)}=-2(N^2-1)-6n^2(N-1).\eqno(4.26 b)
$$

The central charge of the coset model $U(NK)/SU(N)_K \sim  SU(K)_N\times U(1)$
is the sum of the three values $(4.18 b)$, $(4.23 b)$ and $(4.26 b)$. The final
result is
$$
c_{U(NK)/SU(N)_K}= {N(K^2-1)\over N+K}+1\eqno(4.27 b)
$$
independent of the topological charge as we explained at the end of the
previous
subsection. Of course the result $(4.27 b)$ is the usual value given  by the
coset construction\refto{GKO}
$$
c_{U(NK)/SU(N)_K}=c_{U(NK)} -c_{SU(N)_K}.\eqno(4.29 b)
$$

\head{Conclusions}

We have studied a fermionic coset model $G/H$ with subgroup $H$ admitting
non-trivial topology ($\Pi_1(H)\not=0$). The partition function of this theory
corresponds to a sum over the different topological sectors. For each sector we
showed that the partition function can be factorized in three factors: a
partition function for fermions coupled to a non-abelian monopole, a partition
function of a gauged WZW theory and a partition function for a ghost system
minimally coupled to the same monopole field.  We computed the fermionic zero
modes produced by the monopole background and showed the dependence of the
index of the Dirac operator with the topological sector. With the election of
a particular monopole background (a non-abelian gauge field with support at
infinity) the three subsystems become conformal invariant; moreover,
the presence of
the monopole alters the value of the conformal anomaly of each subsystem. This
change
can be computed exactly using an analogy with the Dotsenko and Fateev's
Coulomb Gas approach to CFT. We showed that, for
each subsystem,
the effect of the monopole is to create a background charge in
the direction of the Cartan
subalgebra which modifies the energy momentum tensor of the theory. This
modification has the general form
$$
\Delta T_{z z}\propto \partial_{z} {\bf J}^{N-1}_{total}
$$
where ${\bf J}^{N-1}_{total}$ is the total current in the direction
$T^{N-1}$. However, since in coset models
this current is constrained to zero by the Lagrange
multiplier, the total energy-momentum tensor does not suffer any change
due to the topology. The variations of the conformal anomaly of the three
subsystems adds up to zero. We proved this result explicitly by computing the
conformal central charge for the three subsystems in the general coset
$U(NK)/SU(N)_K$.

Finally let us mention that this result does not imply the independence of the
model on the topology of the gauge field. In fact, we have showed explicitly
that the dependence of the effective action on the monopole charge
affects the conformal dimensions of the primary fields of each subsystem.
Moreover as we mentioned in section
III, the existence of fermionic zero modes impose selection rules over the
vacuum expectation values of the original fermionic fields which consequently
have a strong dependence on the topological charge\refto{Bardakci,ellos,yo}.

\head{Acknowledgments}  I am very grateful with E. Fradkin and F. Schaposnik
for helpful and illuminating discussions. This work was supported in part by
the National Science Foundation through grant No.DMR91-22385 at the University
of Illinois.
\references

\refis{GKO} P.Goddard, A.Kent and D.Olive, \pl 15B, 88, 1985; \cmp 103, 105,
1985.

\refis{coset} K.Bardakci, E.Rabinovici and B.Saring, \np B299, 151, 1988;
A.Polyakov, Lectures at Les Houches Session XLIX 1988, eds. E.Brezin and
J.Zinn Justin, Elsevier, 1989;
D.Cabra, E.Moreno and C.von Reichenbach, \journal Int. Jour. of Mod. Phys. A
, 5, 2313, 1990.

\refis{coset2} K.Gawedzki and A.Kupiainen, \pl 215B, 119, 1988;
D.Karabali, Q-Han Park, H.Schnitzer and Z.Yang, \pl 16B, 307, 1989.

\refis{Bardakci} K.Bardakci and L.Crescimanno, \np 313B, 269, 1989.

\refis{Dotfat} V.Dotsenko and V.Fateev, \np B240, 312, 1984.

\refis{Fujikawa} K.Fujikawa, \prd 25, 2584, 1982.

\refis{FMS} D.Friedan, E.Martinec and S.Shenker, \np B271,  93, 1986.

\refis{Solo} R.Gamboa Saravi, M.Muschietti and J.Solomin, \cmp 93, 407, 1984.

\refis{ellos} M.Manias, C.Naon and M.Trobo, \prd 41,  3174, 1990;
D.Cabra, M.Manias, F.Schaposnik and M.Trobo, \prd 43,  3508, 1991.

\refis{Polyakov} A.Polyakov and P.Wiegmann, \pl 131B, 121, 1983; \pl 141B,
223, 1984.

\refis{jacobian} E.Fradkin, C.Naon and F.Schaposnik. \prd 36, 3809, 1987.

\refis{cuerdistas} E.Verlinde and H.Verlinde, \np B288, 357, 1987;
E.D`Hoker and D.Phong, \rmp 60, 917, 1988.

\refis{Cachoyyo} H.de Vega and F.Schaposnik, \prl 56, 2564, 1986.

\refis{letogus} H.de Vega, \prd 18, 2932, 1978; L.Cugliandolo and G.Lozano,
\prd 39, 3093, 1989.

\refis{nielsen} P.M.Dirac, \journal Ann. Math., 36, 657, 1935; S.Adler,
\prd 6, 3445, 1972; R.Jackiw and C.Rebbi \prd, 14, 517, 1976; N.Nielsen
and B.Schroer, \np, B127 493, 1977.

\refis{cm} D.Cabra and E.Moreno, \journal J. Phys. A, 23, 4711, 1990.

\refis{yo} E.Moreno, to appear in {\it Mod. Phys. Lett. A}.

\refis{wakimoto} M.Wakimoto, \cmp 104, 605, 1986;  B.L.Feigin and E.V.Frenkel,
\journal Usp. Mat. Nauk., 43, 227, 1988;  A.B. Zamolodchikov, unpublished;
A.Gerasimov, A.Morozov, M.Olshanetsky, A.Marshakov and  S.Shatashvili,\journal
Int. Jour. of Mod. Phys. A, 5, 2495, 1989; P.Bouwknegt, J.McCarthy and K.Pilch,
\journal Prog. Theor. Phys. Supp., 102, 67, 1990.

\refis{georgi} H. Georgi, {\it Lie algebras in particle physics}
, Benjamin-Cummings, (1982).

\endreferences

\end